\documentclass{aa} 

\usepackage{natbib}
\bibpunct{(}{)}{;}{a}{}{,} 

\usepackage{graphicx}
\usepackage{txfonts}
\begin{document} 

\title{Steepening of the 820~$\mu$m continuum surface-brightness
  profile signals dust evolution in TW Hya's disk}

\titlerunning{820 $\mu$m continuum signals dust evolution in TW Hya's disk}

\author{Michiel R. Hogerheijde\inst{1}
 \and David Bekkers\inst{1}
 \and Paola Pinilla\inst{1}
 \and Vachail N. Salinas\inst{1}
 \and Mihkel Kama\inst{1}
\and Sean M. Andrews\inst{2}
 \and Chunhua Qi\inst{2}
 \and David J. Wilner\inst{2}}

\institute{Leiden Observatory, Leiden University, PO Box 9513, 2300
 RA, Leiden, The Netherlands\\ \email{michiel@strw.leidenuniv.nl}
\and
Harvard-Smithsonian Center for Astrophysics, 60 Garden Street MS42,
Cambridge, MA 02138, USA
}

\date{Received; accepted}

\abstract
{Grain growth in planet-forming disks is the first step toward the
  formation of planets. The growth of grains and their inward drift
  leaves a distinct imprint on the dust surface-density distribution
  and the resulting surface-brightness profile of the thermal
  continuum emission.}
{We determine the surface-brightness profile of the continuum emission
  using resolved observations at millimeter wavelengths of the disk
  around TW Hya, and infer the signature of dust evolution on the
  surface density and dust opacity.}
{Archival ALMA observations at 820 $\mu$m on baselines up to 410
  k$\lambda$ are compared to parametrized disk models to determine
  the surface-brightness profile.}
{Under the assumption of a constant dust opacity, a broken radial
  power law best describes the dust surface density, with a slope of
  $-0.53\pm 0.01$ from the 4.1~au radius of the (already known) inner
  hole to a turn-over radius of $47.1\pm 0.2$ au, steepening to
  $-8.0\pm 0.1$ at larger radii. The emission drops below the
  detection limit beyond $\sim$60 au.}
{The shape of the dust surface density is consistent with theoretical
  expectations for grain growth, fragmentation, and drift, but its
  total dust content and its turn-over radius are too large for TW
  Hya's age of 8--10 Myr even when taking into account a radially
  varying dust opacity. Higher resolution imaging with ALMA of TW Hya
  and other disks is required to establish if unseen gaps associated
  with, e.g., embedded planets trap grains at large radii or if
  locally enhanced grain growth associated with the CO snow line
  explains the extent of the millimeter-continuum surface brightness
  profile. In the latter case, population studies should reveal a
  correlation between the location of the CO snow line and the extent
  of the millimeter continuum. In the former case, and if CO freeze
  out promotes planet formation, this correlation should extend to the
  location of gaps as well.}


\maketitle


\section{Introduction}\label{s:intro}

Within the core-accretion scenario, formation of planets starts inside
disks around newly formed stars with the coagulation of
(sub)micron-sized dust grains into larger objects \citep[see,
e.g.,][for a review]{lissauer2007}. Evidence for grain growth follows
from the spectral slope at submillimeter-to-centimeter wavelengths
\citep[e.g.,][]{draine2006} and has been found toward many disks
\citep{beckwith1990, andrews2005, rodmann2006, lommen2009,
  ricci2010oph, ricci2010taurus, ricci2011, mann2010, guilloteau2011,
  ubach2012}. As grains grow to millimeter sizes, they start to
decouple dynamically from the gas and, no longer supported by gas
pressure, begin to feel the `head-wind' of the gas and drift inward
\citep{whipple1972,weidenschilling1977}. This results in radially more
compact millimeter continuum emission (roughly probing
millimeter-sized dust) as compared to near-infrared scattered light
(probing submicron-sized dust) and CO (probing the cold gas), as
detected to an increasing number of disks
\citep{panic2009,andrews2012,degregorio2013,walsh2014}. The high
sensitivity of the Atacama Large Millimeter / submillimeter Array
(ALMA) allows detailed investigation of the radial distribution of
millimeter-sized grains and critical comparison to theoretical
expectations of their growth and dynamics.

One of the first disks for which radial migration of millimeter-sized
dust was inferred, is found around the nearby T Tauri star \object{TW
  Hya} \citep{andrews2012}. At a distance of $53.7\pm6.2$ pc
\citep[Hipparcos;][]{vanleeuwen2007}, TW Hya is the closest gas- and
dust-rich planet-forming disk; it is observed close to face-on with an
inclination of $7^\circ\pm1^\circ$
\citep{qi2004,hughes2011,rosenfeld2012}. TW Hya has a mass of
$0.55\pm0.15$ M$_\odot$ and, although some uncertainty exists
\citep{vacca2011}, an estimated age of 8--10 Myr \citep{hoff1998,
  webb1999, delareza2006, debes2013}. Its disk has a gas mass of at
least 0.05 M$_\odot$ as inferred from the emission of HD
\citep{bergin2013}, consistent with a gas-to-dust ratio of 100 and the
derived dust mass of 2--$6\times 10^{-4}$ M$_\odot$
\citep{calvet2002,thi2010}. The outer radius of the disk as seen at
near-infrared wavelengths and in the emission of CO is $\sim$200--280
au \citep{weinberger2002, qi2004, andrews2012, debes2013}. From the
spectral energy distribution (SED) an inner hole in the disk of
$\sim 4$ au was found \citep{calvet2002, hughes2007,
  menu2014}. \citet{andrews2012} show that millimeter-sized grains, as
traced by 870 $\mu$m SMA observations, extend out only to $\sim 60$
au, indicating significant radial inward drift. \citet{menu2014} found
a population of even larger grains at smaller radii from 7~mm JVLA
observations. At near-infrared wavelengths, \citet{debes2013} show a
depression of emission around 80 au, while \citet{akiyama2015}
detected structure in polarized scattered light at radii of 10--20 au
possibly indicating a gap, or changes in the scale height or opacity
of micron-sized grains.

In this paper we re-analyse archival ALMA Cycle 0 data of the 820
$\mu$m (365.5 GHz) continuum emission of TW Hya with an angular
resolution of $\sim 0{\farcs}4$. This observing wavelength is near 870
$\mu$m, the wavelength used by Andrews et al. (2012) to infer the 60
au outer radius for millimeter-sized grains. We do not consider
available ALMA continuum data at 106 GHz and 663 GHz, because we aim
to make a direct comparison to the 870 $\mu$m SMA data without the
added degeneracy between dust surface density and dust emissivity as
function of wavelength. The sensitivity of the archival ALMA data
allows detailed comparison of the continuum visibilities as function
of deprojected baseline length with parametrized models of the
disk. Section \ref{s:data} presents the data and describes how the
visibilities vary with deprojected baseline length. Section
\ref{s:mdlfit} describes our disk model and the fitting methods,
yielding the radial surface-brightness profile of millimeter-sized
grains as well as the underlying dust surface density and
opacity. Section \ref{s:discussion} discusses our results in the light
of theoretical models of dust grain growth, fragmentation, and
transport, and Section \ref{s:summary} summarizes our findings.

\section{Observations and data reduction}\label{s:data}

The ALMA archive contains two Cycle 0 data sets with continuum
observations of TW Hya near 870 $\mu$m. These data sets are
2011.0.00340.S (340.S, for short), covering wavelengths of 804--837
$\mu$m (frequencies of 348--373 GHz), a total bandwidth of 0.94 GHz,
baselines up to 370 m (460 k$\lambda$), and a total on-source
integration time of 5449 s; and 2011.0.00399.S (399.S, for short),
covering 867--898 $\mu$m (334--346 GHz), a total bandwidth of 0.24
GHz, baselines up to 345 m (430 k$\lambda$), and a total on-source
integration time of 4615 s. For both data sets, 3C279 and J0522$-$364
served as bandpass calibrators; Ceres, and for some of data set 399.S,
Titan, as flux calibrators; and J1037$-$295 as gain calibrator. Median
system temperatures were 230--260 K for S.340 and 110--130 K for
S.399. The 340.S data were previously presented by \citet{qi2013}, who
do not analyse the continuum emission in detail.

After verifying the calibration of these data sets, to improve the
complex gain calibration we iteratively performed self-calibration on
the continuum down to solution intervals of 60 s. We use CASA 4.2.1
for the calibration. We reach noise levels of 0.50 mJy~beam$^{-1}$
(340.S) and 0.59 mJy~beam$^{-1}$ (399.S). Uniformly weighted
synthesized beams are $0{\farcs}45\times 0{\farcs}37$ (340.S) and
$0{\farcs}45 \times 0{\farcs}40$ (399.S). Integrated fluxes of 1.7 Jy
(340.S) and 1.3 Jy (399.S) are found.

\begin{figure}
\centering
\includegraphics[width=9cm]{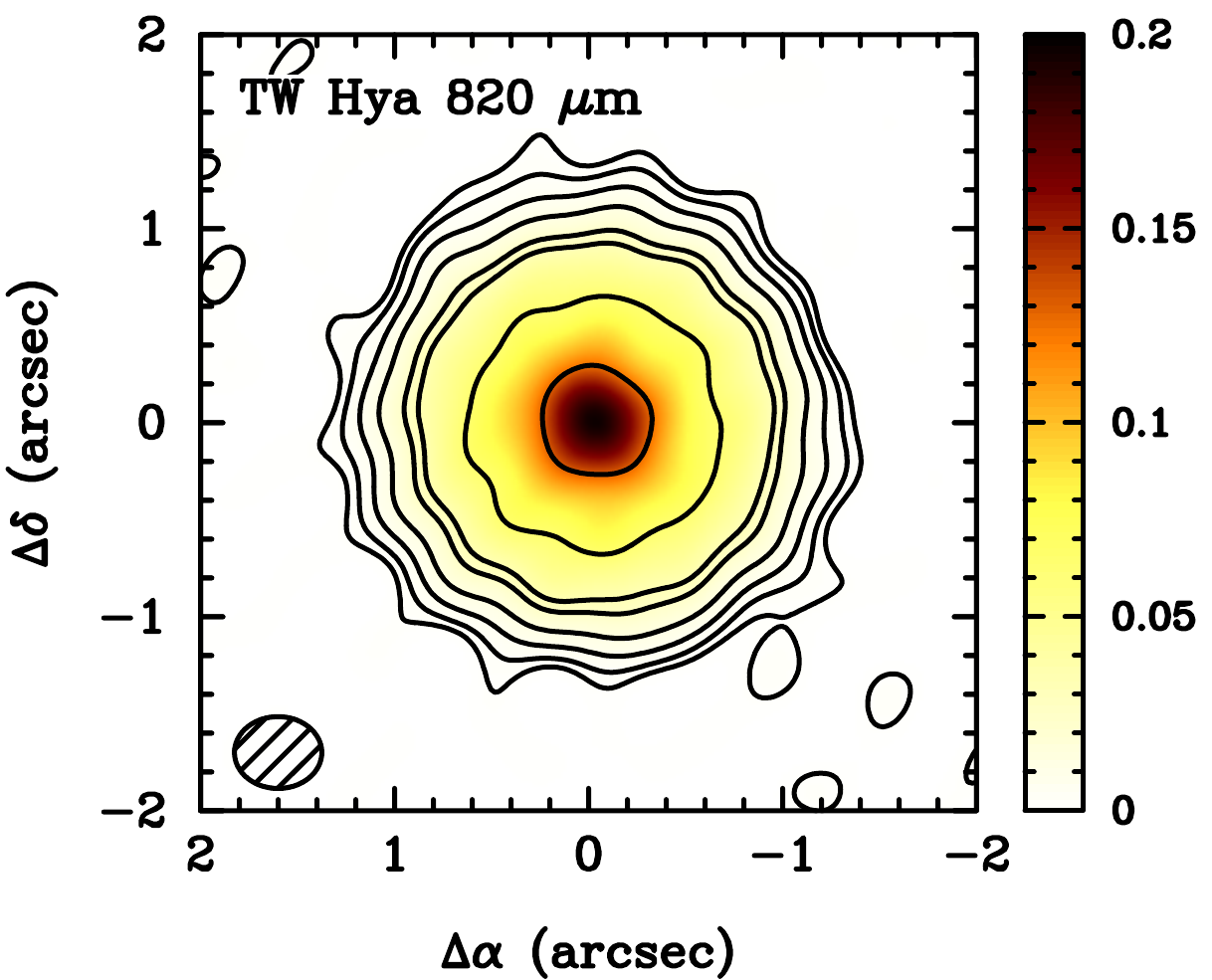}
\caption{Image of the 820 $\mu$m (365.5 GHz) continuum of TW Hya based
  on data set 2011.0.00340.S. We re-imaged these data, and obtain a
  result indistinguishable from \citep{qi2013}.  The image
  construction employed self-calibration and uniform weighting. The
  resulting beam size, indicated in the lower left, is
  $0{\farcs}45\times 0{\farcs}37$ at a position angle of
  $-56^\circ$. The color scale is in units of Jy~beam$^{-1}$ and
  contours are drawn at 1.5, 6.0, 12.0, 24.0, 48.0, 64.0, and
  128. mJy~beam$^{-1}$. The peak intensity is 195.8 mJy~beam$^{-1}$
  and the rms noise level is 0.50 mJy~beam$^{-1}$.}
\label{f:continuum}
\end{figure}

In the remainder of this paper, we focus on data set 340.S which has
slightly higher resolution and lower noise.  Although we can combine
both data sets into a single set at an effective observing wavelength,
with a $\sim 30\%$ lower resulting noise, we instead prefer to use
data set 399.S to independently confirm our results (as is indeed the
case). Other than that, we do not further discuss this data set in
this paper. Figure \ref{f:continuum} shows the continuum image
obtained from data set 340.S.

Adopting a position angle for the (projected) disk's major axis of
$155^\circ$ (east of north) and an inclination of $+6^\circ$
\citep{qi2013, rosenfeld2012}, we derive average continuum
visibilities (real and imaginary part) in 41 10-k$\lambda$ wide radial
bins of deprojected baseline length. We use the equations of
\citet{berger2007} \citep[as referenced in][]{walsh2014} to carry out
the deprojection. Given the near face-on orientation of TW Hya's disk,
the deprojection corrections are small.

\begin{figure}
\centering
\includegraphics[width=9cm]{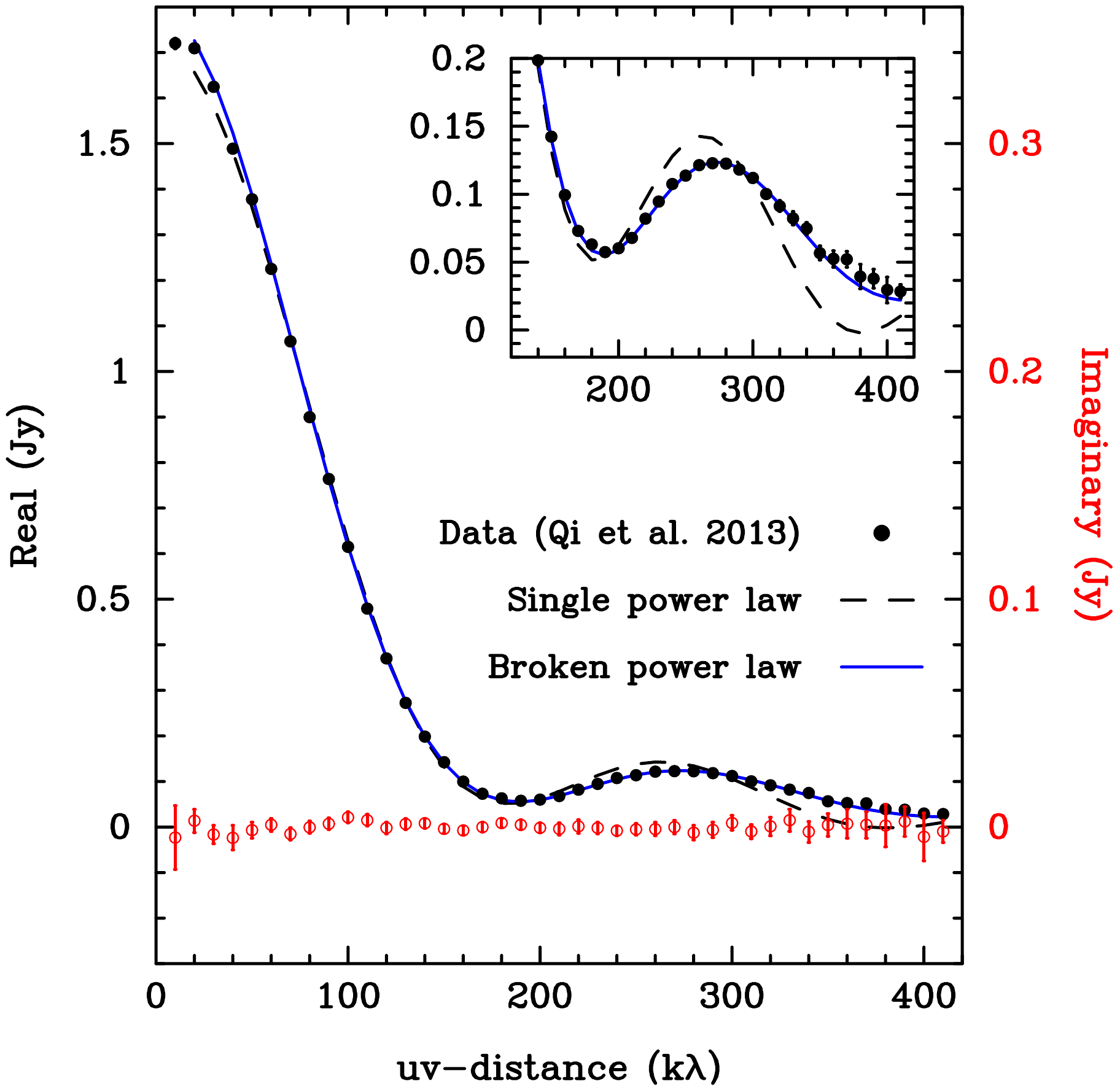}
\caption{Real (black points) and imaginary (red points) parts of the
  visibilities of project 340.S \citep[previously presented
  in][]{qi2013} in Jy vs deprojected baseline length in
  k$\lambda$. The inset shows the region of 120--420 k$\lambda$ in
  greater detail. The dashed black curve shows the recalculated single
  power-law model of \citet{andrews2012} (their model pC), which fails
  to reproduce the data on baselines $>290$ k$\lambda$. A much better
  fit is found for a broken power-law model described in Section
  \ref{s:mdlfit} (solid blue curve). Because both models are
  symmetric, the imaginary parts of their visibilities are all 0 (not
  shown).}
\label{f:visplot}
\end{figure}

Figure \ref{f:visplot} plots the real and imaginary parts of the
continuum visibilities against the deprojected baseline length. The
imaginary parts are all zero to within the accuracy, indicating that
the emission is symmetric around the source center at the resolution
of our observations. Apart from the much higher signal-to-noise, the
real part of the visibilities show a behavior similar to that shown by
\citet{andrews2012}, with a drop off from 1.72 Jy at the shortest
baselines to a minimum of 0.057 Jy near 190 k$\lambda$, followed by a
second maximum of 0.123 Jy around 275 k$\lambda$ and a subsequent
decrease to 0.028 Jy at our longest baseline of 410 k$\lambda$. The
data of \citet{andrews2012} continue to longer baselines of 600
k$\lambda$ with fluxes approaching 0 Jy.

We overplot the data with the model presented by \citet{andrews2012}
(their model pC) which they find to best fit the continuum data. We
recalculate the model at the exact observing wavelengths of the ALMA
data. Our model is geometrically thin and uses only the midplane
temperatures as calculated by \citet{andrews2012} (see Section
3). Our model curve therefore differs in detail from the one shown by
\citet{andrews2012}, but qualitatively shows the same behavior. In
particular, it fits the data well out to baselines of 290 k$\lambda$
but severely under-predicts the emission on baselines of 300 and 400
k$\lambda$, corresponding to angular scales of
$0{\farcs}5$--$0{\farcs}7$ (27--37 au). In the next section we attempt
to find a model description that provides a better fit.

\section{Model fitting}\label{s:mdlfit}

To further investigate the disk structure that best describes the
observed emission, we consider the deprojected and radially binned
real visibilities of data set 340.S. We fit to the 41 data points
plotted in Fig.\ \ref{f:visplot} rather than the full set of (complex)
visibilities to speed up calculations. Since the observed emission is
symmetric (imaginary visibilities are all 0) and the 41 data points in
Fig.\ \ref{f:visplot} fully sample the curve and do not average out
any detected asymmetric structure, this simplification is allowed. As
a consequence of using only the binned visibilities, our fitting
places relatively little weight on the shortest-baselines which
represent the total flux. Because we are primarily interested in the
spatial distribution of the emission, rather than the total amount,
our results are not affected. However, even a small discrepancy in
total flux would be accentuated if we would calcuate a residual
image. For this reason, we refrain from imaging the residuals and
exclusively work with the visibilities.

The real part of the visibilities, $V$, are related to the intensity
distribution of the source by
\begin{equation}
V(r_{uv}) = 2\pi \int_0^\infty I_\nu(R)\, J_0(2\pi Rr_{uv})\, RdR,
\label{e:vis}
\end{equation}
where $r_{uv}$ is the baseline length, $I_\nu(R)$ is the intensity as
function of radius $R$, and $J_0$ the zeroth-order Bessel function of
the first kind \citep[see, e.g.,][]{hughes2007}. We solve the integral
in equation (\ref{e:vis}) by summation using a step size of 0.1 au,
which was found to be sufficiently small.

The intensity $I_\nu(R)$ follows from the expression
\begin{equation}
  I_\nu(R) = B_\nu\bigl(T(R)\bigr)\, \Bigl(1-\exp\bigl(-\tau(R))\Bigr),
\label{e:intens}
\end{equation}
where $B_\nu\bigl(T(R)\bigr)$ is the Planck function, $T(R)$ is the
dust temperature, and $\tau(R)$ is the dust optical depth which is
given by the product of the dust surface density $\Sigma$, the dust
opacity $\kappa$ at the observing wavelength, and $\cos\,i$, where $i$
is the inclination. The factor $\cos\,i$ is very close to unity for TW
Hya.

We follow the midplane temperature as calculated by Andrews et
al. (2012), which reproduces the observed SED. Here, $T(R)$ is
characterized by a very steep drop off in the inner few au, with
$T(R)\propto R^{-q_0}$ with $q_0=2.6$, followed by a very shallow
decrease across the remainder of the disk, with $T(R)\propto R^{-q_1}$
and $q_1=0.26$. Physically, this corresponds to the stellar radiation
being stopped by the high opacity in the inner few au of the disk just
outisde the inner hole, and the remainder of the disk being heated
indirectly by stellar radiation absorbed in the disk surface and
reradiated vertically \citep[e.g.,][]{chiang1997}. In our model, the
temperature at the turnover radius, $R_0$ (7 au), $T_0$, is a free
parameter. In all cases, we find best-fit values for $T_0$ close to
the values found by \citet{andrews2012}. Furthermore, $R_0$ is much
smaller than the resolution of our data, and $T_0$ is degenerate with
the optical depth $\tau(R)$.  Therefore their exact values do not
change our results. The steep rise of the temperature in the inner few
au adds an unresolved component to the continuum emission. This
provides a vertical offset to the real part of the
visibilities. Indeed, the observed real part of the visibilities never
become negative (Fig.\ref{f:visplot}), showing that the sharp increase
in temperature at small radii is essential in our model even if we do
not resolve these scales.

We choose to describe the optical depth $\tau(R)$ as the product of a
constant dust opacity $\kappa_{365.5}$ and a radially varying dust
surface density distribution $\Sigma(R)$. For the dust opacity at the
(average) observing wavelength of 365.5 GHz $\kappa_{365.5}$, we adopt
the same value as employed by Andrews et al. (2012),
$\kappa_{365.5}=3.4$ cm$^2$ g$^{-1}$ (dust). We describe the dust
surface density $\Sigma(R)$ as a (set of) radial power law(s) between
an inner radius $R_{\rm in}$ and an outer radius $R_{\rm out}$. In
Section \ref{s:discussion_drift} we further discuss what happens if
also $\kappa$ varies radially, as would be naturally expected.

Table \ref{t:parameters} lists the fixed and 4 or 6 (depending on the
model) free parameters of our model. Estimates of the parameters
follow from a Markov chain Monte Carlo (MCMC) exploration of the
parameter space, minimizing $\Sigma (x_i - y_i)^2/2\sigma_i^2$, where
$x_i$ and $y_i$ are the 41 observed and modeled binned real
visibilities, and $\sigma_i$ the observed dispersion in each of the
bins. These $\sigma_i$ correspond to the error on the averaged real
visibilities and follow from the dispersion of the data points in the
bin divided by the square root of the number of data points. Any
undetected asymmetric structures also contribute to $\sigma_i$, which
ranges from 2.0 to 12.0 mJy, with a median value of 3.0 mJy. We use
the \texttt{emcee}\footnote{\texttt{http://dan.iel.fm/emcee}}
implementation of the (MCMC) method, which uses affine invariant
ensemble sampling \citep{goodman2010,foreman-mackey2013}. Figure
\ref{f:app_emcee1} in Appendix \ref{app} plots the results.

Assuming a single radial power law for the surface density, the
relevant free parameters in our model are the surface-density
distribution $\Sigma(R)$, the inner hole radius $R_{\rm in}$, and the
outer radius $R_{\rm out}$. In all cases we find inner radii of the
order of a few au, consistent with SED estimates of the inner
hole. The best-fit model of \citet{andrews2012} (their model pC) has
$\Sigma(R)\propto R^{-0.75}$ and a surface density of $\Sigma_0$ of
0.39 g cm$^{-2}$ at 10 au. Keeping this $\Sigma_0$ fixed and using a
single power power-law for the radial dependence
($\Sigma\propto R^{-{p_0}}$), we find very similar values as
\citet{andrews2012} for the slope ($p_0=0.70\pm 0.01$ vs 0.75) and
outer radius ($R_{\rm out}=55.7\pm 0.1$ au vs 60 au; see Table
\ref{t:parameters}). The quoted errors on the parameters are formal
fitting errors, only meaningful within the model assumptions. The
emission in Fig. \ref{f:continuum} is convolved with the synthesized
beam and therefore appears to extend beyond the value for
$R_{\rm out}$ found from the visibilities (57 au, corresponding to
$1{\farcs}1$). Figure \ref{f:visplot} shows the model curve
corresponding to the single-power fit. This curve is similar (but not
identical) to the curve for model pC in \citet{andrews2012}. Although
the parameters are nearly identical, our model is vertically
isothermal, unlike model pC, resulting in a slightly different
curve. Regardless, both models have equal problems in reproducing the
visibilities on baselines of 300--400 k$\lambda$.

\begin{table}
\caption{Model Parameters}
\label{t:parameters}
\centering
\begin{tabular}{lcc}
\hline\hline
Parameter & Single & Broken \\
 & power law & power law \\
\hline 
\multicolumn{3}{c}{Fixed parameters:}\\
\hline 
$\Sigma_{\rm10\,au}$ (g cm$^{-2}$) & $ 0.39$ & $ 0.39$ \\
$\kappa_{365.5}$ (cm$^2$ g$^{-1}$) & $ 3.4$ & $ 3.4$ \\
$q_0$ & $ 2.6$ & $ 2.6$ \\
$q_1$ & $ 0.26$ & $ 0.26$ \\
$R_0$ (au) & $ 7.0$ & $ 7.0$ \\
\hline
\multicolumn{3}{c}{Free parameters:}\\
\hline
$R_{\rm in}$ (au)    & $4.44\pm 0.02$  & $4.07 \pm 0.03 $\\
$R_{\rm out}$ (au)   & $55.7\pm 0.1$   & $200 \pm 55$\\
$p_0$               & $0.70 \pm 0.01$ & $0.53 \pm 0.01 $\\
$p_1$               &                 & $8.0 \pm 0.1$\\
$T_0$ (K)           & $28.4\pm 0.2$   & $26.7 \pm 0.2 $\\
$R_{\rm break}$ (au) &                  & $47.1 \pm 0.2 $\\
\hline
\end{tabular}
\tablefoot{Quoted uncertainties are formal fitting errors only, and do not
  include the effect of the model assumptions or the distance to TW Hya. }
\end{table}

After this single radial power law as a description for the surface
density, the next simplest model consists of a broken power law, with
$\Sigma(R) \propto R^{-p_0}$ inside a radius $R_{\rm break}$ and
$\Sigma(R) \propto R^{-p_1}$ outside this radius. This adds two free
parameters to the model $p_1$ and $R_{\rm break}$. A very good fit to
the data is found (Fig. \ref{f:visplot}, Table \ref{t:parameters} and
Fig. \ref{f:app_emcee2}). In this best-fit model, the radial drop off
of the surface density is initially modest ($p_0=0.53\pm 0.01$) out to
a radius of $47.1\pm 0.2$ au, followed by a very sharp drop at larger
radii ($p_1=8.0\pm 0.1$). Given the fast drop of the surface density,
no useful constraint on the outer radius is found ($200\pm 55$ au), as
the emission drops below the detection limit. This fit is
significantly better than the single power law fit, with respective
$\chi^2$ values of 1437 and 283. A likelihood-ratio test indicates
that the broken power law model describes the data better than the
single power law model with a probability lager than 0.999.

The broken power law solution found above should not be confused with
the exponentially tapered models that are often used \citep[see, for
example,][]{hughes2008}. Such models are described by a surface
density $\Sigma=\Sigma_c (R/R_c)^{-\gamma}\exp[-(R/R_c)^{2-\gamma}]$.
\citet{andrews2012} already show that exponentially tapered models do
not fit the millimeter continuum visibilities of TW Hya: there are no
values of $\gamma$ that give a sufficiently shallow drop off of
$\Sigma$ at small $R$ and a sufficiently steep fall off at large
$R$. Only our broken power law can combine these slopes.

In this analysis, we have parameterized the emission with a constant
dust opacity $\kappa$ and a radially varying surface density
$\Sigma(R)$. We stress that our real constraints are on the dust
optical depth $\tau(R)=\kappa\Sigma$. This optical depth follows the
same radial broken power law found above. The functional description
of the surface-brightness profile is more complex, since following
equation (\ref{e:intens}) it also includes the slope of the
temperature via the Planck function and the slope of the optical depth
via the factor $(1-\exp(-\tau))$. The optical depth drops from 2.1 at
4.1 au to 0.57 at 47 au, making the emission moderately optically
thick and preventing us from taking the limit for small $\tau$,
$\lim_{\tau\to 0} (1-\exp(-\tau))=\tau$.

\section{Discussion}\label{s:discussion}

\subsection{Dust drift}\label{s:discussion_drift}

The broken power-law distribution obtained for the dust surface
density resembles predictions from \citet{birnstiel2014} who consider
the effects of radial drift and gas drag on the growing grains. At
`late' time ($\sim$ 0.5--1 Myr in their models), the dust surface
density follows a slope of approximately $-1$ out to $\sim 50$ au,
after which it steepens to approximately $-10$. Our inferred surface
density profile, although different in detail, shows a similar
steeping, further strengthening the suggestion of \citet{andrews2012}
that the grains in TW Hya's disk have undergone significant growth and
drift. However, \citet{birnstiel2014} consider the total dust surface
density while our fit is essentially to the 820 $\mu$m contiunuum
optical-depth profile. This is dominated by grains of roughly 0.1--10
mm \citep{draine2006} but grains of all sizes contribute. As noted by
\citet{birnstiel2014}, the total dust surface density consist of a
radially varying population of dust at different sizes, with larger
grains having more compact distributions than smaller grains. This is
certainly the case for TW Hya, given that grains small enough to
scatter near-infrared light extend out to 200--280 au
\citep{debes2013} while 7~mm observations by \citet{menu2014} show
emission concentrated toward only the inner several au. From
observations at a single wavelength, we cannot infer both the dust
surface density and the dust opacity as function of radius.

Various authors describe simulations of grain evolution including
growth, fragmentation, and drift \citep[e.g., see][for a
review]{testi2014}.  \citet{pinilla2015} follow the description of
\citet{birnstiel2010} to describe the advection-diffusion differential
equation for the dust surface density, simultaneously with the growth,
fragmentation, and erosion of dust grains via grain-grain
collisions. We do not perform a detailed simulation of the TW Hya disk
here, since this would require knowledge of the gas surface denisty
distribution and comparison to multi-wavelength continuum
observations. Instead, we note that any simulation of realistic disks
using the prescriptions of \citet{pinilla2015} results in a very low
820 $\mu$m continuum surface brightness and in radial profiles that
have turnover radii $R_{\rm break}\lesssim 20$ au at ages $>$5 Myr,
and therefore cannot explain the observations for the 8--10 Myr of TW
Hya.  This age vs surface brightness (extent) problem was already
noted before \citep[e.g.,][]{pinilla2012,pinilla2015,testi2014}. The
conundrum of TW Hya's surface-brightness profile is that it resembles
\emph{in shape} the signatures of grain evolution and drift while in
\emph{in flux and size} it is too bright and extended for these same
mechanisms to have operated over the 8--10 Myr lifetime of the disk.

One possible explanation is that TW Hya's disk was originally much
more massive and larger than the 200--280 au now seen. However, to
have a turn-over radius of $\sim$ 50 au, after 10 Myr, requires an
initial disk size of 800--1000 au and a mechanism to remove all CO and
small dust outside 200 au over the past 10 Myr to match current
observations of CO line emission and near-infrared scattered light. We
do not consider this a likely scenario.

\subsection{The presence of unseen gaps}\label{discussion_gaps}

Another possibility to explain the apparent youth of TW Hya's disk
(both in terms of the turn-over radius of the surface-brightness
distribution of the millimeter continuum and its significant gas and
dust content), lies in comparison to the recent high resolution
imaging results of millimeter continuum emission of HL Tau
\citep{alma_brogan2015}. HL Tau is a very young disk ($<$1 Myr), but
its millimeter continuum image shows a remarkable set of bright and
dark rings. The presence of these rings have been variously
interpreted as the result of gap clearing by embedded planets
\citep{dipierro2015,dong2015,pinte2015}, secular gravitational
instabilities \citep{youdin2011, takahashi2014}, or
emissivity-depressions due to accelerated grain growth to
decimeter-sizes at the location of the frost lines of dominant ices
and clathrates \citep{zhang2015,okuzumi2015}.

Millimeter-sized grains are known to become `trapped' outside gaps in
so-called transitional disks because of the reversed pressure gradient
\citep[e.g.,][]{vandermarel2013, casassus2013}. Therefore, one or more
embedded planets inside the TW Hya disk could have `trapped'
millimeter-sized grains and prevented further inward migration. This
mechanism has been invoked by \citet{pinilla2012}, who show that even
small pressure maxima are sufficient to slow down the inward migration
of millimeter-sized dust. 

We investigated if the currently available ALMA data place constraints
on the presence of gaps inside the TW Hya disk (other than the known
gap inside $\sim 4$ au). Starting with the two-slope model, we include
a gap centered on 20 au characterized by a Gaussian width and a
depth. We choose the value of 20 au to be near the center of the
extent of the millimeter-sized grains. We then repeat the procedure of
the previous section to find the best-fit parameters for a series of
gap widths and depths. By optimizing the other free model parameters
we allow these to maximally `compensate' for the effect of the
presence of the gap on the visibilities. We find that we can exclude,
as defined by an increase of the $\chi^2$ value by a factor of two or
more, models with gaps that are entirely empty with FWHM widths of
more than 6.6 au and models with gaps that have only a depth of 80\%
to widths of 16.6 au. Assuming a gap width of 5 Hill radii, such gaps
correspond to planet masses of 0.7--10 M$_{\rm Jup}$
\citep{dodson-robinson2011}. For comparison, the widths and depths of
the gaps in HL Tau are well within the allowed range. We conclude that
TW Hya's disk may very well contain planet-induced gaps as detected
toward HL Tau, which are capable of halting further inward migration
of grains and explain the observed millimeter-continuum
surface-brightness profile. 

\subsection{Dust temperature and the CO snow line}\label{discussion_temp}

The discussion above attributes the characteristics of the
distribution of the millimeter-wave continuum to the surface density
distribution and opacity of the grains. According to equation
(\ref{e:intens}) the explanation could equally well lie with the dust
temperature. At radii inside the turn-over of the emission, up to 47
au, the disk is known to be cold, $<20$K, as witnessed by the CO
freeze out probed by N$_2$H$^+$ emission \citep{qi2013}. A shallow
surface-density distribution combined with a steep temperature drop
off outside 47 au cannot explain the observations, since this requires
unrealistically low temperatures ($\ll$10 K). A steeper density
distribution combined with a temperature rise around 30--40 au followed
by a return to nominal temperatures can fit the observations, but
requires temperatures as high as 70 K, clearly inconsistent with the
detection of N$_2$H$^+$ that would be rapidly destroyed by gas-phase
CO at such temperatures.

The N$_2$H$^+$ data from \citet{qi2013} place the CO snow line at 30
au. Just outside this snow line, grain growth may receive a boost by
the accumulation of frost resulting from mixing of material across the
snow line. This cold-finger effect was earlier described by
\citet{meijerink2009} and invoked by \citet{zhang2015} to explain the
dark rings in the HL Tau millimeter-continuum image as locations of
growth to decimeter-sized particles at the location of frost
lines. Similarly in TW Hya, the location of the turn-over radius in
the millimeter emission may be determined, not by the inward migration
of millimeter-sized grains, but by a local enhancement of growth of
(sub) micron-sized grains to millimeter sized grains in the 30--47 au
region outside the CO snow line. If the efficiency of grain growth is
locally sufficiently enhanced, this may increase the dust opacity at
820 $\mu$m and explain the observed surface-brightness profile. Zhang
et al. (in prep.) present an analysis of four protoplanetary disks
with high resolution continuum observations, and find that regions
just beyond CO snowlines show locally enhanced continuum emission.
Detailed modeling of grain growth and drift in the presence of ices
and the resulting dust opacity is required to further explore this
mechanism. If this mechanism is indeed effective, population studies
of disks should reveal a correlation between the location of the CO
snow lines\footnote{In the case of a disk with a temperature structure
  that varies with time, the \emph{effective} (or time-averaged)
  location of the CO snow line is what matters here.} and the sizes of
the millimeter continuum emission disk.

\section{Summary}\label{s:summary}

We analyzed archival ALMA Cycle 0 365.5 GHz continuum data with
baselines up to 410 k$\lambda$ and conclude the following.

\begin{enumerate}

\item{The interferometric visibilities are described by a disk with a
    broken power law radial surface-desnity distribution and contstant
    dust opacity, with slopes of $-0.53\pm 0.01$ and $-8.0\pm 0.2$ and
    a turn-over radius of $47.1\pm 0.2$ au. The outer radius is
    unconstrained, as the emission drops below the detection limit
    beyond $\sim$57 au.}

\item{Under the assumption of a constant dust opacity $\kappa$, the
    corresponding surface-density distribution resembles in shape the
    one expected in the presence of grain growth and radial migration
    for `late times' as calculated by \citet{birnstiel2014}.}

\item{The total surface brightness and the turn-over radius of 47 au
    are too large for a source age of 8--10 Myr. This implies that
    either the disk formed with a much larger size and mass, that one
    or more unseen embedded planets have opened gaps and halted inward
    drift of millimeter-sized grains, or that the region outside the
    CO snow line at radii of 30--47 au has a significantly locally
    enhanced production rate of millimeter-sized grains boosting the
    dust opacity at 820 $\mu$m.}

\end{enumerate}

Higher resolution observations with ALMA are essential to explore the
reason for the location of the turnover at 47 au (see, e.g., Zhang et
al. in prep.). Such observations can easily show the presence of one
or more gaps, as shown by the HL Tau image. Near-infrared observations
\citep{debes2013, akiyama2015, rapson2015} suggest the presence of
structures at, respectively, 80 au, 10--20 au, and $\sim 23$ au that
are consistent with (partially filled) gaps. Whether these correspond
to unresolved structures of millimeter-wave emitting grains is unknown
but likely.  If, in spite of this, no gaps in the millimeter emission
are found, high resolution multi-wavelength observations of the
brightness distribution can be contrasted to models of grain growth
and migration in the presence of the CO snow line, to explore if an
increase in grain growth in the 30--47 au region can be sufficiently
efficient to create the observed millimeter-continuum brightness
profile. There is even the possibility that both scenarios operate:
increased grain growth outside the CO snow line has led to the
formation of a planet that subsequently has halted further migration
of millimeter-sized grains. If so, the correlation mentioned above
between CO snow-line locations and the sizes of the millimeter
continuum emission disks, should extend to the locations of gaps as
well.

\begin{acknowledgements}
  This paper makes use of the following ALMA data:
  ADS/JAO.ALMA\#2011.0.00340.S and ADS/JAO.ALMA\#2011.0.00399.S. ALMA
  is a partnership of ESO (representing its member states), NSF (USA)
  and NINS (Japan), together with NRC (Canada), NSC and ASIAA
  (Taiwan), and KASI (Republic of Korea), in cooperation with the
  Republic of Chile. The Joint ALMA Observatory is operated by ESO,
  AUI/NRAO and NAOJ.  The research of MRH and VNS is supported by
  grants from the Netherlands Organization for Scientific Research
  (NWO) and the Netherlands Research School for Astronomy (NOVA). This
  work made use of PyAstronomy.
\end{acknowledgements}



\begin{appendix}
\section{Results of the \texttt{emcee} model fitting}
\label{app}

\begin{figure*}
\centering
\includegraphics[width=14cm]{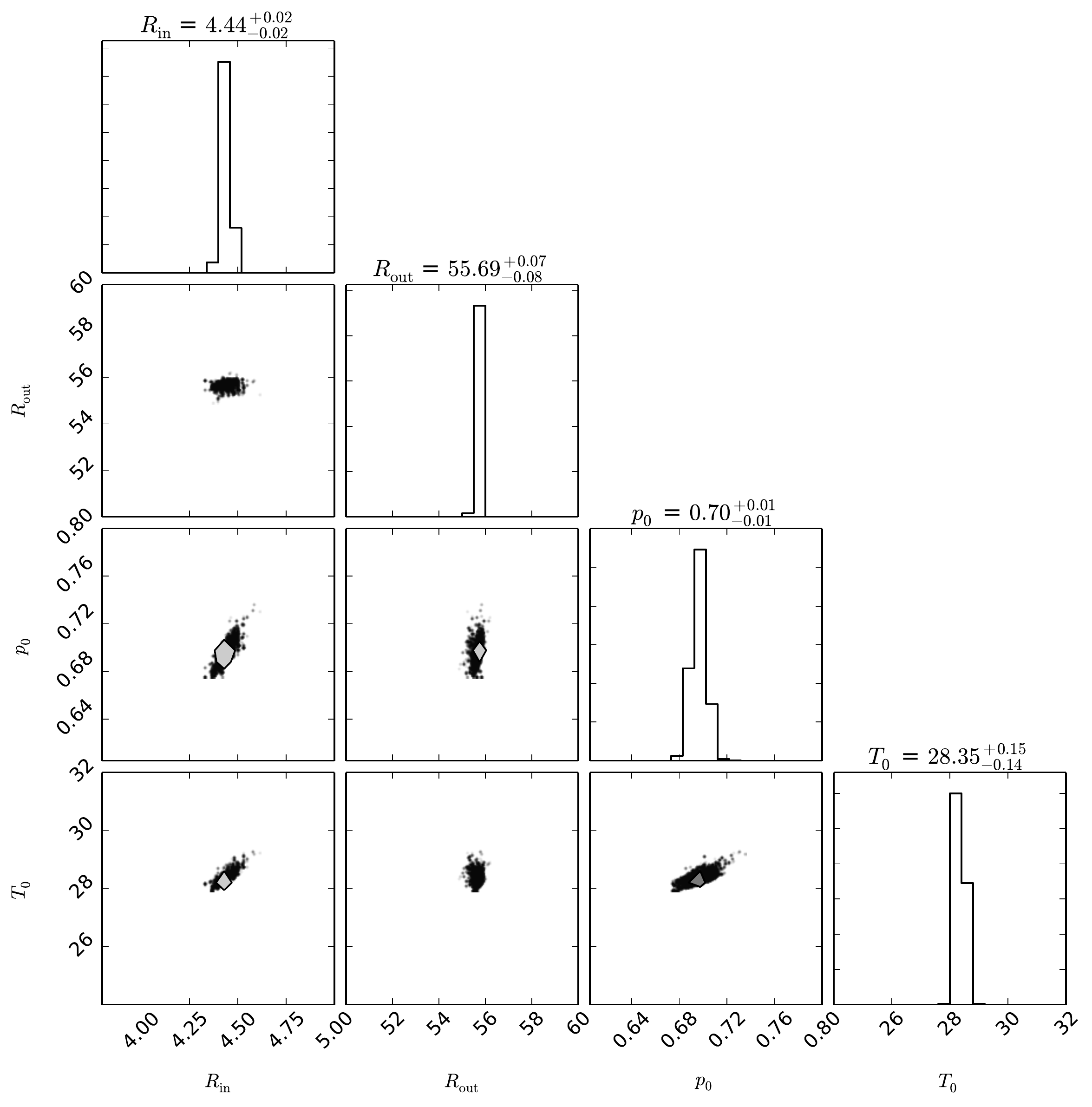}
\caption{Results of the \texttt{emcee} MCMC optimization using a
  single radial power law for the surface density. The panels show
  one- and two-dimensional projections of the posterior probability
  functions of the free model parameters. The panels along the
  diagonal show the marginalized distribution of each of the
  parameters as histograms; the other panels show the marginalized
  two-dimensional distributions for each set of two parameters.
  Contours (barely visible in the tightly constrained probability
  distributions) and associated greyscale show 0.5, 1, 1.5 and
  2$\sigma$ levels. Best-fit values and error estimates are listed
  above the panels along the diagonal.}
\label{f:app_emcee1}
\end{figure*}

\begin{figure*}
\centering
\includegraphics[width=18cm]{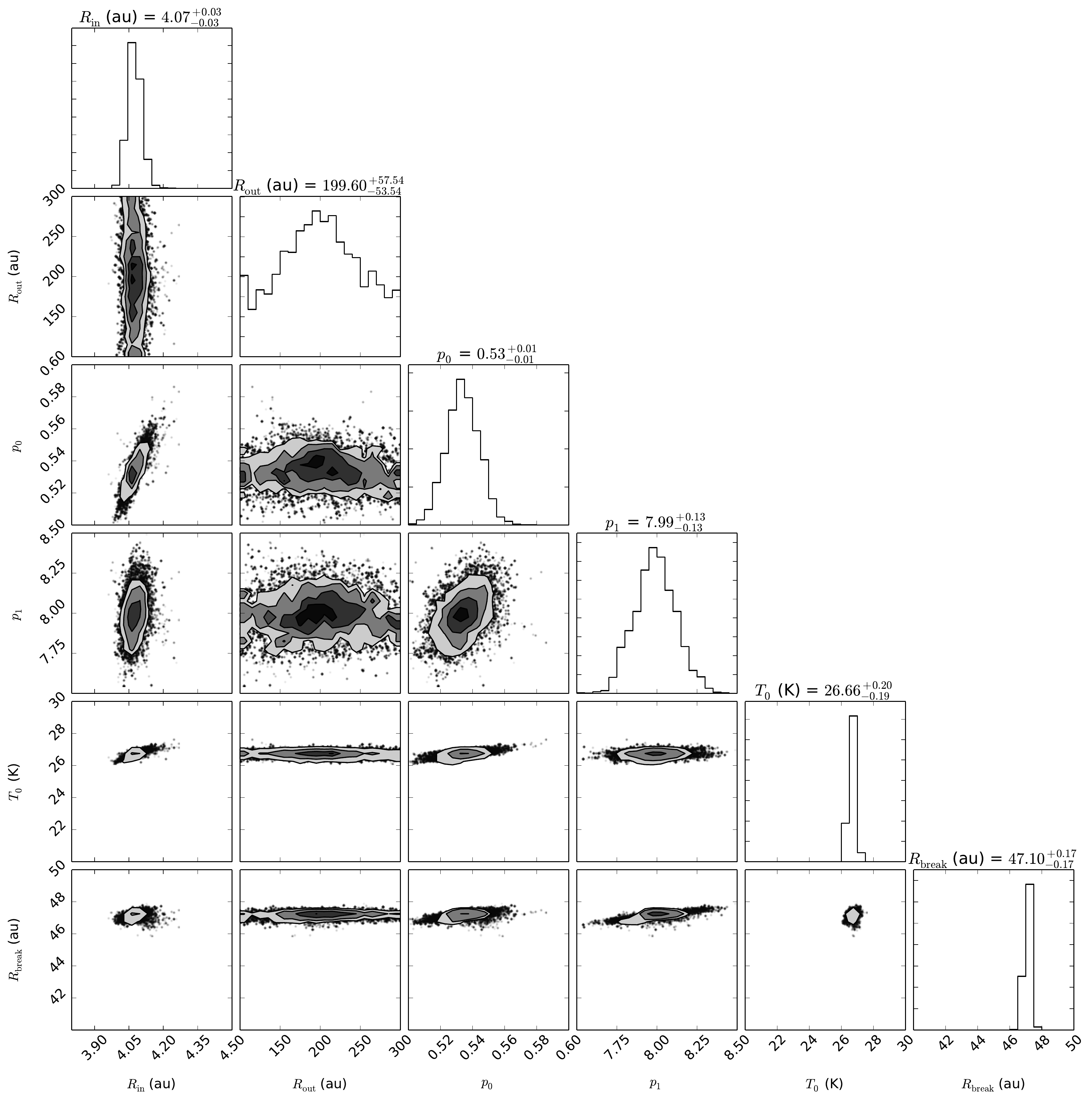}
\caption{Results of the \texttt{emcee} MCMC optimization using a
  broken radial power law for the surface density. Otherwise, the
  figure is similar to \ref{f:app_emcee1}. No useful constraints on
  the outer radius $R_{\rm out}$ are obtained.}
\label{f:app_emcee2}
\end{figure*}

\end{appendix}

\end{document}